\documentclass{article}

\usepackage{amsmath,amssymb}
\usepackage[super]{cite}

\begin{document}

\title{Remarks on the Origins of Path Integration:\\
Einstein and Feynman\thanks{%
To appear in: Proceedings of the International Conference `Path Integrals - New
Trends and Perspectives,' Dresden, Germany, 23--28 September 2007.}}

\author{T. Sauer\\
\footnotesize{Einstein Papers Project}\\[-0.10cm]
\footnotesize{California Institute of Technology 20-7}\\[-0.1cm]
\footnotesize{Pasadena, CA 91125, USA}\\[-0.10cm]
\footnotesize{E-mail: tilman@einstein.caltech.edu}}

\date{}

\maketitle

\begin{abstract}
I offer some historical comments about the 
origins of Feynman's path integral approach, as an alternative approach to
standard quantum mechanics. Looking at the interaction between Einstein and
Feynman, which was mediated by Feynman's thesis supervisor John Wheeler,
it is argued 
that, contrary to what one might expect, the significance of the interaction
between Einstein and Feynman pertained to a critique of classical field
theory, rather than to a direct critique of quantum mechanics itself. 
Nevertheless, the critical perspective on classical field theory became
a motivation and point of departure for Feynman's space-time approach
to non-relativistic quantum mechanics.\\[0.5cm]
\small{Keywords: History of quantum mechanics, Einstein, Feynman.}
\end{abstract}



\section{Introduction}

In this paper, I am interested in the genesis of Feynman's path integral approach
to non-relativistic quantum mechanics. I take Feynman's 1948 paper on ``A Space-Time
Approach to Quantum Mechanics''\cite{Feynman1948} as the point in time when
the approach was fully formulated and published and made available to the community
of physicists. I will take a look into the prehistory of Feynman's 1948 paper.
I shall not
attempt to give anything like a balanced, or even complete historical account
of this prehistory. Instead, I will focus on a little footnote in Feynman's paper:
\begin{quote}
The theory of electromagnetism described by J.A.~Wheeler and R.P. Feynman, Rev.\ Mod.\ 
Phys.\ {\bf 17}, 157 (1945) can be expressed in a principle of least action 
involving the coordinates of particles alone. It was an attempt to quantize this theory, 
without reference to the fields, which led the author to study the formulation of
quantum mechanics given here. The extension of the ideas to cover the case of more general
action functions was developed in his Ph.D.\ thesis, ``The principle of least action
in quantum mechanics'' submitted to Princeton University, 1942. \cite[p.~385]{Feynman1948}
\end{quote}
My guide in organizing my remarks will be to look at what we know about any direct
and indirect interaction between Feynman and
Einstein. Let me briefly motivate this focus on Einstein and Feynman.

Feynman was born in New York in 1918, 
did his undergraduate studies at MIT, and took his Ph.D.\ with John A.~Wheeler at Princeton
University in 1942, before going to Los Alamos during the war years. After the war,
he was first at Cornell and in 1951 he went to Caltech.
In 1954, Feynman received the Einstein Award,\cite{nyt} as a 36-year old man
for his work on quantum electrodynamics that in 1965 would earn him 
the Nobel prize for physics.

The Einstein Award was a prestigious award, established in 1949 in Einstein's
honor, but it seems that Einstein had not much to do with the awarding of the prize to
Feynman. 

At the time of Feynman's receiving the Einstein award, Einstein himself was a 
76-year old world famous man. He had been living in Princeton since 
his emigration from Nazi-Germany in 1933 and was 
scientifically engaged in a search for a unified field theory of gravitation
and electromagnetism.\cite{Sauer2007a} But he also still thought about problems of the 
foundations of quantum mechanics.
Among his extensive research notes and manuscripts with calculations along the
unfied field theory program, there is, e.g., a manuscript page from around 1954 with
a concise formulation of Einstein's of the famous Einstein-Podolsky-Rosen incompleteness
argument for standard quantum mechanics. Probably in reaction to David Bohm's
reformulation of the original argument, Einstein here also formulates the
incompleteness argument for spin observables.\cite{Sauer2007b}

With both Feynman and Einstein being concerned with the foundations of 
quantum mechanics, one might hope that an interaction
between the two physicists, if there was any, might give us some insight
into the historical development of our understanding of the principles of
quantum theory.

A similar question was also asked once by Wheeler. In 1989, after Feynman's
death, he recalled:
\begin{quote}
Visiting Einstein one day, I could not resist telling him about Feynman's new way
to express quantum theory.\cite{Wheeler1989}
\end{quote}
After explaining the basic ideas of Feynman's path integral approach to Einstein,
Wheeler recalls to have asked:
\begin{quote}
``Doesn't this marvelous discovery make you willing to
accept quantum theory, Professor Einstein?'' He replied in a serious voice,
``I still cannot believe that God plays dice. But maybe,'' he smiled, ``I
have earned the right to make my mistakes.''\cite{Wheeler1989}
\end{quote}
So is this the end of my story?

According to Feynman's own account, he himself met Einstein only twice. One 
of these encounters was at the occasion of  
Feynman's first technical talk, as a young graduate student, in the
Princeton physics department. The occasion probably took place in late 1940. 
Wheeler had suggested that 
Feynman was to talk on their joint work, and Feynman recalls
\begin{quote}
Professor Wigner was in charge of the colloquium, so after I said I would do
it, he told me that he had heard from Wheeler about the work and he knew
something about it. I think we had discussed it a little bit with him. And
he thought it was important enough that he had taken the liberty to invite especially 
Professor Henry Norris Russell from the astronomy department, the great astronomer,
you know, John von Neumann from the mathematics department, the world's great mathematician,
and Professor Pauli, who was visiting from Zurich, would be there. And Professor
Einstein had been especially invited---and although he never comes to the colloquia,
he thinks he will come!\\
So I went through fire on my first. I must have turned a yellowish-green or
something [...].%
\footnote{\cite[p.~133]{Interview1966}. I am quoting from the transcript of an oral history
interview conducted by C.~Weiner with Feynman in 1966. For slightly different versions of 
the episode, see also \cite[pp.~64f]{joking} and \cite[p.~66]{Leighton2006}.}
\end{quote}
Feynman continues to recount details of this seminar, he relates how his excitement and
anxiety abated once he started to talk about physics, and indicates how some members of 
his audience, including Einstein, reacted to his presentation in question time.

In the following, I will take this encounter between Feynman and Einstein as a
point of departure for a historical argument: the interaction between Feynman and Einstein
reminds us of a significant historical context of discovery of the path integral method.
This original context is still prominently visible in Feynman's 1942 thesis\cite{Brown2005} 
but it is already reduced to a footnote in his 1948 publication.\cite{Feynman1948} 
For an appreciation of the path-integral
method, even today, it may nevertheless still be useful to recall the historical 
circumstances of its discovery.

Specifically, I will address and discuss the following four questions:

\begin{enumerate}
\item What is the Wheeler-Feynman theory that Feynman presented in his first 
seminar at Princeton?
\item What does Einstein have to do with this?
\item What does this have to do with path integrals?
\item Why is this context of the origin of the path integral approach
only mentioned in a footnote in Feynman's 1948 paper?
\end{enumerate}
  
Most of the information that the argument
is based on can be found in Schweber's book on the 
history of quantum electrodynamics\cite{Schweber1994}. The significance of the 
Wheeler-Feynman theory for Feynman's subsequent development is also emphasized
by Feynman himself in his Nobel lecture.\cite{Feynman1966}

\section{The Wheeler-Feynman absorber theory}

The results of their joint work that Feynman presented in the Princeton physics 
colloquium were not published at the time. Feynman gave his presentation again,
shortly thereafter, at a meeting of the American Physical Society
in Cambridge, Massachussetts, that took place
on 21 and 22 February 1941. Of this talk, an abstract was published.\cite{FW1941}
There exists also a typescript by Feynman giving an account of the theory,\cite{FP6.1}
dated to spring 1941 \cite[p.~383]{Schweber1994}.
The abstract identifies radiative damping as a problem in Lorentz's classical electron
theory and in Dirac's theory of a point electron, and then summarizes the main points
of Feynman's paper:
\begin{quote}
We postulate (1) that an accelerated point charge in otherwise free space does
{\it not} radiate energy; (2) that, in general, the fields which act on a given
particle arise only from {\it other} particles; (3) that these fields are 
represented by one-half the retarded plus one-half the advanced Li\'enard-Wiechert
solutions of Maxwell's equations. In a universe in which all light is eventually
absorbed, the absorbing material scatters back to an accelerated charge a field,
part of which is found to be independent of the properties of the material.
This part is equivalent to one-half the retarded field {\it minus} one-half
the advanced field generated by the charge. It produces radiative damping
(Dirac's expression) and combines with the field of the source to give
retarded effects alone.\cite{FW1941} 
\end{quote}

A detailed account of the Wheeler-Feynman theory was published after the war in two papers.
The first paper appeared in 1945 under the title `Interaction with the Absorber
as the Mechanism of Radiation.'\cite{WF1945}
As indicated in a first footnote to the title, this paper essentially gives an account of the
theory that Feynman had presented in 1941. 

In a long second footnote to the title of the paper, Wheeler then explains that 
this first paper was actually planned to be the third part of a projected series 
of five papers.
Of the missing four parts, only the second one actually appeared, four years
later, in 1949, under the title `Classical Electrodynamics in Terms of Direct
Interparticle Action'.\cite{WF1949}

The core of the Wheeler-Feynman theory
thus concerned a special problem that arose
out of a broader research program, laid out, in part, in the later 1949 paper.
In qualitative terms, the broader research program concerned this.

Among the many difficulties with attempts to come to a 
quantum theory of electrodynamics in the late thirties, Wheeler and Feynman thought some 
had to do with difficulties that occur already at the level of classical electrodynamic field
theory. As a radical 
response, Wheeler and Feynman questioned whether the notion
of an electromagnetic field is, in fact, a useful one. They argued that one should
in principle be able to express all electromagnetic phenomena in terms of direct interaction
between point-like particles. Any notion of a field would be a derived concept. The primary
notion would be a collection of point-like charges that interact with each other through
Li\'enard-Wiechert retarded and advanced potentials. 

They found that
such a theory is expressible in terms of an action principle that involves a variation
over the world-lines of charged electrons.  
They wrote the action principle as\cite{WF1949}
\begin{eqnarray}
J = -&\sum_am_ac\int (-da_{\mu}da^{\mu})^{\frac{1}{2}}+\sum_{a < b}(e_ae_b/c)\notag\\
&\times \iint \delta(ab_{\mu}ab^{\mu})(da_{\nu}db^{\nu}) = \text{extremum},
\label{eq:J}
\end{eqnarray}
where the sums are over electrons of mass $m_a$ and charge $e_a$, $da$ denotes derivative
with respect to the respective proper time, and $ab^{\mu}\equiv a^{\mu}-b^{\mu}$ is short 
for the four-vector of the separation between the particles, a somewhat unusual notation
introduced in order to be able to make use of the Einstein summation convention. 
The attractive feature of this action is that ``all of mechanics and electrodynamics is
contained in this single variational principle.''\cite[p.~425]{WF1949}
Note that the single action principle incorporates both the Maxwell equations and
the Lorentz force law.  The idea and the action (\ref{eq:J}) were known before, they
can be found, in more or less explicit terms, in older papers
by Schwarzschild\cite{Schwarzschild1903}, Tetrode\cite{Tetrode1922}, and
Fokker\cite{Fokker1929}.

The only problem with this formulation was the issue of radiative reaction.
In classical theory,
an accelerated electron radiates and loses energy to the field. To avoid the
notion of a field, Wheeler and Feynman postulated
that a single electron alone in the universe, if accelerated, would, in fact, {\it not}
radiate. Instead, they succeeded to show that radiative reaction can arise in a
universe with a surrounding material that absorbs all outgoing radiation.
The electrons of the absorber interact with the electron at the source through
advanced potentials, such that an accelerated electron feels a radiative force.
This is the main point that Feynman was elaborating on in his Princeton seminar.

\section{Einstein and the electromagnetic arrow of time}

Feynman recalled that immediately
after his presentation, Pauli asked critical questions and then asked Einstein whether
he would agree.

\begin{quote}
Anyway, Professor Pauli got up immediately after the lecture. He was sitting next to
Einstein. And he says, ``I do not think this theory can be right because of this, that
and the other thing---'' it's too bad that I cannot remember what, because the theory is
not right, and the gentleman may well have hit the nail on the bazeeto, but I don't know,
unfortunately, what he said. I guess I was too nervous to listen, and didn't understand
the objections. ``Don't you agree, Professor Einstein?'' Pauli said at the end of his
criticism. ``I don't believe this is right---don't you agree, Professor Einstein?''\\
Einstein said, ``No,'' in a soft German voice that sounded very pleasant to me, and said 
that he felt that the one idea, the one thing that seemed to him, was that the principles
of action and distance which were involved here were inconsistent with the field views,
the theory of gravitation, of general relativity. But after all general relativity is not
so well established as electrodynamics, and with this prospect I would not use that as
an argument against you, because maybe we can develop a different way of doing 
gravitational interaction, too.\\
Very nice. Very interesting. I remember that.%
\footnote{\cite[p.~134]{Interview1966}, see also \cite[p.~66]{joking}, 
	\cite[pp.~67--68]{Leighton2006}.}
\end{quote}

We also know, both from Feynman \cite[p.~133]{Interview1966} and
Wheeler\cite[p.~167]{Wheeler1998} as well as, independently, from a letter by Wheeler to 
Einstein, that Feynman and Wheeler visited Einstein once in his house in Princeton
and discussed the 
``interpretation of the force of radiation in terms of advanced and retarded action at
a distance.''\cite{AEA23442}
It is unclear when the meeting took place,%
\footnote{It is even unclear whether the meeting in Einstein's house took
place before or after Feynman's Princeton colloquium. In 1966, Feynman did not remember
but was ``pretty sure'' that it was before the colloquium ``because he knew me,'' 
see \cite[pp.~133,139]{Interview1966}. Wheeler \cite[p.~167]{Wheeler1998} recalls
that it was ``while working on our second action-at-a-distance paper'', but from his letter
to Einstein, we know that it must have been before November 1943, see
also \cite[p.~118]{Gleick1992}.}
and I am not aware of any detailed account of the discussion that took place, 
but it seems that Einstein
alerted Feynman and Wheeler to existing literature on the subject, including
some in which he himself was involved.
In a footnote to their 1945 paper, Wheeler and Feynman acknowledge Einstein's
input:
\begin{quote}
We are indebted to Professor Einstein for bringing to our attention the ideas of Tetrode
and also of Ritz, [...].\cite[n.~10]{WF1945}
\end{quote}
Somewhere else in the article, they 
\begin{quote}
recall an inconclusive but illuminating discussion carried on by Ritz and Einstein
in 1909, in which ``Ritz treats the limitation to retarded potentials as one of the
foundations of the second law of thermodynamics while Einstein believes that the
irreversibility of radiation depends exclusively on considerations of
probability.''\cite[p.~160]{WF1945}
\end{quote}
The Einstein-Ritz controversy\cite{Ritz1908,Einstein1909,RitzEinstein1909},
from which they quoted,
was about the origin of irreversibility of electromagnetic
radiation phenomena.\cite{EarmanPreprint}
In the 1941 typescript, Feynman observed that their theory is in full agreement with
Einstein's position against Ritz, that the fundamental electrodynamical equations
are time-reversal invariant, and that the radiative irreversibility is a macroscopic, 
statistical phenomenon:
\begin{quote}
The apparent irreversibility in a closed system, then, either from our point of view
or the point of view of Lorentz is a purely macroscopic irreversibility. The 
present authors believe that all physical phenomena are microscopically
reversible, and that, therefore, all apparently irreversible phenomena are solely
macroscopically irreversible. (\cite[p.~13.1]{FP6.1}; quoted in \cite[p.~386]{Schweber1994}.)
\end{quote}
Feynman here has a footnote saying
\begin{quote}
That this and the following statement are true in the Lorentz theory was
emphasized by Einstein in a discussion with Ritz. (Einstein and Ritz,
Phys.\ Zeits.\ 10, p323, (1909)). Our viewpoint on the matter discussed
is essentially that of Einstein. (We should like to thank Prof.~W.~Pauli
for calling our attention to this discussion.) (ibid.)
\end{quote}

Although Pauli is credited here for alerting Feynman to the Ritz-Einstein 
controversy, we may assume that the point was also a topic when
Feynman and Wheeler discussed their ideas with Einstein during their
visit at his Princeton home. There is, in any case, an English translation,
in Feynman's hand, of the Ritz-Einstein controversy\cite{RitzEinstein1909} 
in the Feynman papers.\cite{RitzEinsteinRPF}

\section{Path integrals for actions with no Hamiltonian}

In 1942, Feynman was recruited for the Los Alamos project. Before leaving
for Los Alamos, Wheeler urged Feynman to write up his thesis.\cite{WF1942}
Feynman's thesis\cite{Brown2005}
is not directly dealing with the Wheeler-Feynman absorber theory but it rather
gives a discussion of the `Principle of Least Action in Quantum Mechanics', and is,
in fact, a direct forerunner of Feynman's 1948 paper. But the thesis is
very explicit about its original motivation. The discussion of quantizing systems
expressed in terms of a Lagrangian is given in the context
of solving the general problem of finding a quantum version of the 
Wheeler-Feynman theory of action-at-a-distance.
The main point here is that
\begin{quote}
the theory of action at a distance finds its most natural expression in a principle
of least action, which is of such a nature that no Hamiltonian may be derived from it. 
That is to say the equations of motion of the particles cannot be put into Hamiltonian form
in a simple way. This is essentially because the motion of one particle at one time 
depends on what another particle is doing at some other time, since the interactions are
not instantaneous.\cite{ThesisDraft}%
\end{quote}
This is not just a remark made in (a draft version of) the preface to motivate the approach.
An example that derives directly from the action (\ref{eq:J}) is discussed also 
in the body of the text. At some point, Feynman explains how to generalize the
quantization procedure to more general actions, for example those involving
time-displaced interactions:
\begin{quote}
The obvious suggestion is, then, to replace this exponent by $\frac{i}{\hbar}$
times the more general action. The action must of course first be expressed
in an approximate way in terms of $q_i$, $t_i$ in such a way that as the
subdivision becomes finer and finer it more nearly approaches the action
expressed as a functional of $q(t)$.\\
In order to get a clearer idea of what this will lead to, let us choose a simple 
action function to keep in mind, for which no Hamiltonian exists. We may take,
\[\mathcal{A}=\int_{-\infty}^{\infty}\left\{\frac{m\dot{x}(t)^2}{2} 
- V(x(t)) + k^2\dot{x}(t)\dot{x}(t+\tau)\right\}dt,\]
which is an approximate action function for a particle in a potential $V(x)$ and
which also interacts with itself in a mirror by half advanced and half retarded
waves, [...].\cite[p.~41]{Brown2005}
\end{quote}
In the 1941 typescript Feynman comes close to showing how this simple
action follows from the general action (\ref{eq:J}) by considering the special
case of two charges at a distance apart in otherwise free space, neglecting their
electrostatic interaction.
Of course, the path integral quantization of actions that are non-local in time
is considerably more involved\cite{Schulman1995}
and Feynman does not give an explicit discussion of
his example. Nevertheless, it confirms his remark in the (actual) preface of the thesis 
which
\begin{quote}
is concerned with the problem of finding a quantum mechanical description
applicable to systems which in their classical analogue are expressible  by
a principle of least action, and not necessarily  by Hamiltonian equations of
motion.\cite[p.~6]{Brown2005}
\end{quote}

\section{The demise of the early context of path integration}

In 1949, even before the second of the 
Wheeler-Feynman papers appeared in print, Feynman himself submitted another one of 
his famous papers, entitled
`Space-Time Approach to Quantum Electrodynamics.'\cite{Feynman1949} In it,
one finds this little footnote:
\begin{quote}
These considerations make it appear unlikely that the contention of J.A.~Wheeler
and R.P.~Feynman, Rev.\ Mod.\ Phys.\ {\bf 17}, 157 (1945), that electrons do not
act on themselves, will be a successful concept in quantum electrodynamics.
\cite[p.~773]{Feynman1949}
\end{quote}
Why did Feynman retract a basic assumption of his joint work with Wheeler, with explicit
reference to their earlier paper?
Two years later, Feynman wrote a letter to Wheeler asking him about his opinion about
the status of their earlier work:

\begin{quote}
I wanted to know what your opinion was about our old theory of action at a distance.
It was based on two assumptions:\\
(1) Electrons act only on other electrons;\\
(2) They do so with the mean of retarded and advanced potentials.\\
The second proposition may be correct but I wish to deny the correctness of the
first. The evidence is two-fold. First there is the Lamb shift in hydrogen
which is supposedly due to the self-action of the electron. [...]\\
The second argument involves the idea that the positrons are electrons going
backwards in time. [...]\\
So I think we guessed wrong in 1941. Do you agree?\cite{FW1951}
\end{quote}
I am not aware of an explicit response by Wheeler to this letter, but several 
remarks in his autobiography\cite{Wheeler1998} indicate that he, too, eventually gave up his
belief in an action-at-a-distance electrodynamics: ``[...] until the early 1950s,
I was in the grip of the idea that Everything is Particles.'' \cite[p.~63]{Wheeler1998}

For Feynman, one of the two reasons for
giving up the theory of action-at-a-distance was an experimental finding, the Lamb shift.
Lamb had presented data from his experiments on the fine structure of hydrogen
at the Shelter Island conference. This conference, devoted to problems of the quantum
mechanics of the electron, took place in June 1947 and was an event of considerable
impact in the history of post-war physics \cite[ch.~4]{Schweber1994}. 
It brought together the leading theorists for the first time after the war for a
meeting which helped to determine the course of American physics in the atomic age. 
9 of the 23 participants ended up being awarded the Nobel prize, a significant
fraction of the participants were of the young generation. 
It was at this conference that Feynman presented his `space-time approach to quantum
mechanics', essentially the work of his thesis, and
soon after the conference
he penned his classic 1948 paper.\cite{Feynman1948}

Incidentally, the Shelter Island conference could have provided an occasion for a third
encounter between Feynman and Einstein: following a suggestion of Wheeler, who was present
as well, Einstein was among the invitees but he declined, due to ill health 
\cite[pp.~169f]{Schweber1994}.
It is tempting to speculate how Einstein would have reacted to Feynman's presentation 
of his new approach to quantum mechanics at this meeting.

When Feynman wrote up his approach for publication, he decided to mention the original
motivation for his work only in passing.
Given the generality of the path integral formulation, 
it may be seen a wise decision on Feynman's part to reduce the historical
context of its genesis to a footnote.

	 
\end{document}